\documentclass[amsmath,amssymb,aip,jcp,10pt,twocolumn]{revtex4-1}

\usepackage{tablefootnote}
\usepackage{dcolumn}
\usepackage{graphicx}

\begin{document}

\title{At what chain length do unbranched alkanes prefer folded conformations?}

\author{Jason N. Byrd}
\email{byrdja@chem.ufl.edu}
\affiliation{Quantum Theory Project, University of Florida, Gainesville, FL 32611}
\author{Rodney J. Bartlett}
\affiliation{Quantum Theory Project, University of Florida, Gainesville, FL 32611}
\author{John A. Montgomery, Jr.}
\affiliation{Department of Physics, University of Connecticut, Storrs, CT 06269}


\begin{abstract}
Short unbranched alkanes are known to prefer linear conformations, while long unbranched alkanes are folded.  It is not known with certainty at what chain length the linear conformation is no longer the global minimum.  To clarify this point, we use {\it ab initio} and density functional methods to compute the relative energies of the linear and hairpin alkane conformers for increasing chain lengths.  Extensive electronic structure calculations are performed to obtain optimized geometries, harmonic frequencies and accurate single point energies for the selected alkane conformers from octane through octadecane.  Benchmark CCSD(T)/cc-pVTZ single point calculations are performed for chains through tetradecane, while approximate methods are required for the longer chains up to octadecane.  Using frozen natural orbitals to unambiguously truncate the virtual orbital space, we are able to compute composite CCSD FNO(T) single point energies for all the chain lengths.  This approximate composite method has significant computational savings compared to full CCSD(T) while retaining $\sim0.15$ kcal/mol accuracy compared to the benchmark results.  More approximate dual-basis resolution-of-the-identity double-hybrid DFT calculations are also performed and shown to have reasonable $0.2-0.4$ kcal/mol errors compared with our benchmark values.  After including contributions from temperature dependent internal energy shifts, we find the preference for folded conformations to lie between hexadecane and octadecane, in excellent agreement with recent experiments [L\"{u}ttschwager, N.  O.; Wassermann, T. N.; Mata, R. A.; Suhm, M. A. {\it Angew. Chem. Int. Ed.} 2013, 52, 463].  
\end{abstract}


\maketitle

\section{Introduction}

Unbranched alkane chains (C$_n$H$_{2n+2}$) are of fundamental importance in
organic chemistry.  They are constituents of fossil fuels and polymers, as well
as important structural motifs in lipids and other biomolecules.  It is clearly
important to understand their conformational and thermochemical properties.  

\begin{figure}[b]
\includegraphics[width=8.26cm]{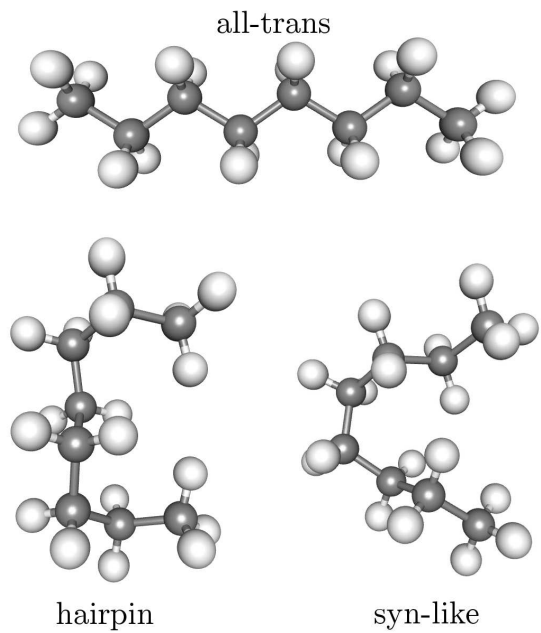}
\caption{\label{diagram}Illustrative optimized alkane structures.}
\end{figure}

The conformer potential energy surface of an unbranched alkane is characterized
by torsional twists which lead from linear chains to highly deformed structures
dominated by intramolecular dispersion forces.  At temperatures less than $300$
Kelvin, short alkanes ($n=4-8$) in the gas phase are well known to prefer the
linear all-trans ($T=180^\circ$, $X=90^\circ$ and $G=60^\circ$ for trans, cross and
gauche dihedral angles respectively) conformation.  However, as the length of
the alkane grows there must be a point where the attractive intramolecular
interactions will cause the chain to self-solvate into a folded conformer.  A
cross-gauche-cross rotation combination 
($T\ldots XGX \ldots T$) 
is sufficient to fold the chain,\cite{grimme2007}
but this creates an energetically unfavorable {\it syn}-pentane like 
conformation.  In addition, the chain ends are not parallel in
this conformation, reducing the possible stabilization due to van der Waals
attraction.  A hairpin conformation with four gauche rotations\cite{goodman1997}
($T\ldots GGTGG\ldots T$) minimizes the number of strained bonds and allows an
energetically favorable parallel arrangement of the chain ends, leading it to be
the suggested global minimum for longer
alkanes.\cite{goodman1997,thomas2006,luttschwager2013}  These three
conformational structures are illustrated in Figure \ref{diagram}.

It is well known that the computation of relative energies involving weak
interactions presents a significant challenge for computational studies.  In the case of
short alkanes ($n=4-6$), standard DFT methods are largely inadequate to describe
the conformer energies and ordering of states\cite{gruzman2009} while {\it ab
initio} methods past second order perturbation theory are necessary for a full
description of the more configurationally complicated transition state conformer
structures.\cite{martin2013}  As the length of the alkane chain increases so
does the importance of a proper treatment of dispersion.\cite{grimme2007}  In the case of octane
($n=8$) second order perturbation theory calculations underestimate the ``bowl"
conformer ($GGTGG$) energy difference while coupled cluster theory (CCSD) will
overestimate the energy difference by $\sim0.5$ kcal/mol compared to the ``gold
standard" inclusion of perturbative triples obtained using CCSD(T).

With these computational difficulties in mind, it is clear why the hairpin
conformer critical chain length (at which it becomes the global minimum) is
difficult to determine accurately. 
Early work by Goodman\cite{goodman1997} using force field and
semi-empirical calculations suggested a turning point anywhere between $n=12$ to
$n=26$, with subsequent force field calculations\cite{thomas2006} also pointing
towards $n=18$ as the critical chain length.  This problem was experimentally
addressed in the recent work of L\"{u}ttschwager {\it et al.}\cite{luttschwager2013}
Their experiment used Raman spectroscopy with a supersonic jet
expansion apparatus and concluded that the critical chain length is between $n=16$
and $n=18$ at temperatures of 100K.  The accompanying theoretical work uses a
local coupled cluster approach to also suggest a critical chain length of $n=18$.

In this paper we will first obtain benchmark {\it ab initio} structures,
electronic and harmonic vibrational energies for the linear (all-trans) and hairpin
alkane conformers of increasing length starting with octane ($n=8$) through
tetradecane ($n=14$).  These benchmark values will be used to characterize
various approximate methods that are extendable to longer chains ($n>14$) with
which we predict the critical alkane chain length.
Although they are not the focus of this study, entropic effects
become important as the temperature increases, and this is discussed briefly
in the conclusions.

\section{Electronic Structure Calculations}

We perform {\it ab initio} and DFT electronic structure calculations on
the n-alkane ($n=8,10-18$) linear and hairpin conformers using
the GAMESS\cite{gamess1993,gamess2005} and ACES III\cite{acesiii2008} quantum
chemistry packages running on the University of Florida HPC, HiPerGator and
University of Connecticut BECAT clusters.  All calculations in this work 
use Dunning's correlation consistent family of basis sets
(cc-pVnZ, n=T,Q).\cite{dunning1989}  For resolution-of-the-identity calculations
the triple-zeta fitting basis set of Weigend {\it et al.}\cite{weigend2002}
(cc-pVTZ-RI) are used in conjunction with the standard cc-pVTZ basis set.
Unless explicitly stated all correlation calculations in this work assume the
frozen-core approximation where all $1s$ carbon orbitals are dropped from the
correlation space while the corresponding last virtual orbital is retained.

For shorter ($n=2-7$) alkane chains it has been
shown\cite{klauda2005,gruzman2009,martin2013} that the quality of the final
optimized geometry is more strongly dependent on the level of the correlation theory
than on the basis set.  Additional tests show that including correlation beyond
that of second order M{\o}ller-Plesset perturbation theory (MP2) is unnecessary.
For the longer alkane species we find that the use of a small basis set (such
as the Pople split valence 6-311G* basis set) that has insufficient polarization
functions will lead to erroneous hairpin structures.  Therefore
geometry optimizations of the alkane conformers used in this work are obtained
using the MP2 level of theory (with analytic gradients) and the cc-pVTZ basis set.  
Harmonic zero point energy (ZPE) shifts are computed numerically using analytic
first derivatives at the MP2/cc-pVTZ all-electron level of theory.  Due
to the cost of numerical MP2 hessians, the massively parallel ACES III program
is used to perform the necessary first derivatives with the caveat that only
all-electron MP2 gradients are available.  This change of theory between
geometries and hessian is found to introduce a negligible error of $~0.03$
kcal/mol when considering relative conformer ZPE shifts.

High level single point energy calculations are computed using coupled cluster
theory with singles, doubles, and perturbative
triples\cite{purvis1982,raghavachari1989,bartlett2007} (CCSD(T)) and the
cc-pVTZ basis set.  Higher order effects such as contributions from the
core-valence correlation energy or higher order excitations (full triples,
quadruples etc.) are not included as their effects are small and cancel nearly
identically in conformational energy differences. 
It is convenient to analyze the calculated energies by orders of
perturbation theory.  In this way the MP2 correlation energy is given by
$\Delta {\rm MP2}=E({\rm MP2})-E({\rm SCF})$, 
while higher order contributions can be
conveniently given as 
$\Delta {\rm CCSD}=E({\rm CCSD})-E({\rm MP2})$,
$\Delta {\rm CCSD(T)}=E({\rm CCSD(T)})-E({\rm MP2})$, 
and $\Delta {\rm (T)}=E({\rm CCSD(T)})-E({\rm CCSD})$.  
Because the basis set convergence of the post-MP2
correlation is much faster than the second order contribution\cite{platts2013},
we can estimate the effects of going to the complete basis set (CBS) limit by
combining large basis MP2 energies and small basis coupled cluster correlation
energies.  Using the cc-pVTZ and cc-pVQZ basis sets, we separately extrapolate
the SCF and MP2 correlation energy using the linear extrapolation formulas of
Schwenke\cite{schwenke2005} 
\begin{equation}
E_\infty({\rm SCF})=E_{n-1}({\rm SCF})+F_{n-1,n}(E_n({\rm SCF})-E_{n-1}({\rm SCF}))
\end{equation}
(with $F_{3,4}=1.3071269$) and Helgaker {\it et al.}\cite{helgaker1997}  
\begin{equation}\label{cbs}
\Delta_\infty {\rm MP2}=
\frac{n^3 \Delta_n {\rm MP2} - (n-1)^3 \Delta_{n-1} {\rm MP2}}{n^3-(n-1)^3} 
\end{equation} 
respectively.  Here $E_n({\rm SCF})$ and $\Delta_n {\rm MP2}$ refers to the SCF and MP2
correlation energy computed with the cc-pVnZ basis set.  Adding in the coupled cluster
correlation energy to form a composite CBS energy
\begin{equation}\label{energy} 
E({\rm CCSD(T)})/{\rm CBS}=E_\infty({\rm SCF})+\Delta_\infty {\rm MP2} +
\Delta {\rm CCSD(T)}
\end{equation}
we obtain a $~0.2$ kcal/mol shift compared to the CCSD(T)/cc-pVTZ relative
energy for the $n=8-14$ alkanes (see Table \ref{benchE}).  As the cost of doing
an MP2/cc-pVQZ calculation for the alkane chains longer than $n=14$ starts to
become prohibitively expensive (over 1800 basis functions and 130 electrons) we
instead omit the CBS correction and include a $0.2$ kcal/mol error estimate
in our final result.

\begin{table}[t]
\caption{\label{benchE}{\rm Benchmark {\it ab initio} conformer energy differences
(hairpin $-$ linear, in kcal/mol)
for the $n=8-14$ alkane chains.  Harmonic ZPE energies were computed at
the MP2/cc-pVTZ level of theory.}}
\begin{tabular}{lrrrr}
   Method
 & C$_{8}$H$_{18}$
 & C$_{10}$H$_{22}$
 & C$_{12}$H$_{26}$
 & C$_{14}$H$_{30}$ \\
\hline
 & \multicolumn{4}{c}{cc-pVTZ} \\
SCF      & 5.19 & 5.50 & 5.88 & 9.76 \\
$\Delta{\rm MP2}$  & -4.10 & -4.76 & -5.59 & -10.47 \\
$\Delta{\rm CCSD}$ & 0.83 & 1.00 & 1.18 & 2.11 \\
$\Delta{\rm (T)}$  & -0.49 & -0.58 & -0.70 & -1.41 \\
ZPE                & 0.58 & 0.60 & 0.58 & 0.69 \\
 & \multicolumn{4}{c}{cc-pVQZ} \\
SCF      & 5.24 & 5.56 & 5.95 & 9.89 \\
$\Delta{\rm MP2}$  & -4.07 & -4.70 & -5.52 & -10.50 \\
 & \\
CCSD(T)  & 1.44 & 1.18 & 0.79 & 0.03 \\
CCSD(T)/CBS & 1.57 & 1.36 & 0.99 & 0.14 \\
\end{tabular}\end{table}

Using the massively parallel ACES III program allowed us to compute the CCSD
energies for all the alkane chain lengths considered here.  However the
$O(o^3v^4)$ (for $o$ occupied and $v$ virtual orbitals) scaling of the
perturbative triples quickly becomes problematic.  With our available
computational resources the perturbative triples contribution could be computed
only for alkane chains up to $n=14$.  In order to alleviate the cost of
including triples it is possible to truncate the virtual space by some amount
$p$, providing a $p^4$ prefactor which can enable larger calculations to be
done.  To do so systematically and unambiguously we use the frozen natural
orbital (FNO) method\cite{taube2005,taube2008} which uses the MP2 density matrix
to make new virtual orbitals.  The Hartree-Fock virtual orbitals can then
be replaced with the appropriately transformed MP2 virtual natural orbitals,
resulting in a set of virtual orbitals sorted by their contribution to the
correlation energy.  The virtual space can then be truncated by examining the
MP2 virtual occupation numbers (eigenvalues of the MP2 density matrix) and
dropping orbitals with an occupation smaller than some predetermined threshold.
In this work we take a threshold of $1\times 10^{-4}$, which results in $~40\%$
of the virtual orbitals being dropped (a prefactor $p^4$ of $0.13$).  Additional
savings are also realized in the form of memory storage and data communication
requirements. 

There are a variety of ways to get a final FNO coupled cluster energy.
Two composite energy schemes are considered:
\begin{equation}\label{fnoa}
E({\rm FNO\; CCSD(T)}) = E({\rm MP2})+\Delta_{\rm FNO} {\rm CCSD(T)}
\end{equation} 
and
\begin{equation}\label{fnob}
E({\rm CCSD\; FNO(T)}) = E({\rm CCSD})+\Delta_{\rm FNO} {\rm (T)}.
\end{equation} 
both of which have comparable accuracies for smaller systems (see 
Table \ref{fnoE}). 
We choose to use the latter composite method
(Equation \ref{fnob}), which makes approximations only in the triples calculation.
Our particular implementation in GAMESS and ACES III uses the converged
$T_1$ and $T_2$ amplitudes from an $\Delta_{\rm FNO}$CCSD calculation to compute the
perturbative triples contribution.  The error associated with using these
amplitudes compared to complete virtual space CCSD amplitudes which are then
truncated by the FNO prescription is small.\cite{deprince2012}
Similar composite methods have been used by DePrince and
Sherrill with comparable accuracy obtained.\cite{deprince2012}

\begin{table*}
\caption{\label{fnoE}{\rm Computed {\it ab initio}
conformer energy differences (hairpin $-$ linear, in kcal/mol)
for the $n=8-18$ alkane chains using the
cc-pVTZ basis set.  Both FNO composite energies are shown, as well as the
temperature dependent enthalpy changes.}}
\begin{tabular}{lrrrrrr}
Method
 & C$_{8}$H$_{18}$
 & C$_{10}$H$_{22}$
 & C$_{12}$H$_{26}$
 & C$_{14}$H$_{30}$ 
 & C$_{16}$H$_{34}$ 
 & C$_{18}$H$_{38}$ \\
\hline
SCF      & 5.19 & 5.50 & 5.88 & 9.76 & 10.11 & 11.06\\
$\Delta{\rm MP2}$  & -4.10 & -4.76 & -5.59 & -10.47 & -11.70 & -13.77 \\
$\Delta{\rm CCSD}$ & 0.83 &  1.00 &  1.18 &   2.11 &   2.34 &   2.75 \\
\\
$\Delta_{\rm FNO}$MP2  & -3.80 & -4.32 & -5.09 & -9.62 & -10.71 & -12.61 \\
$\Delta_{\rm FNO}$CCSD & -0.13 &  0.94 &  1.16 &  2.05 &  2.27 &   2.66 \\
$\Delta_{\rm FNO}$(T)  & -0.59 & -0.49 & -0.59 & -1.23 & -1.39 &  -1.64 \\
 & \\
CCSD FNO(T) & 1.51 & 1.25 & 0.89 & 0.18 & -0.63 & -1.60 \\
FNO CCSD(T) & 1.50 & 1.20 & 0.86 & 0.12 & -0.70 & -1.69 \\
 \\
$\Delta H_{0K}$   & 2.09 & 1.85 & 1.47 & 0.86 &  0.05 & -0.92 \\
$\Delta H_{100K}$ & 2.02 & 1.76 & 1.37 & 0.67 & -0.14 & -1.11 \\
$\Delta H_{298K}$ & 1.78 & 1.51 & 1.11 & 0.35 & -0.46 & -1.43 \\
\end{tabular}
\end{table*}

An approximate theoretical method that scales better than the usual CCSD
$O(o^2v^4)$ calculation with very reasonable accuracy for short
alkanes\cite{gruzman2009} is double-hybrid density functional
theory\cite{grimme2006a,schwabe2008} (DH-DFT).  This method mixes SCF and DFT
exchange with DFT and MP2 correlation energy then corrects the
dispersion energy empirically using Grimme's D3 correction.\cite{grimme2010}  To
facilitate calculations of even larger molecules, we have recently
implemented\cite{byrd2013-b} in GAMESS the
dual-basis\cite{wolinsky2003,liang2004,nakajima2006,steele2006a,steele2009}
SCF method.  Here the SCF energy is approximated by a converged small (truncated)
basis energy calculation.  The large (with polarization functions) basis
contribution is then approximated by a single new Fock matrix constructed from
the projected small basis density matrix.  This approximate SCF method is much
faster than a full SCF calculation with errors comparable to standard
density-fitted SCF methods.  We evaluate the DH-DFT method in GAMESS using
dual-basis DFT married with the resolution-of-the-identity\cite{katouda2009} MP2
method (referred together with the dual-basis SCF to as DB-RI), a further
approximation\cite{steele2006b} that adds trivial errors while shifting all of the leading
computational cost to the single large basis Fock matrix build.  

\section{Computational Results and Discussion}

When computing relative conformer energies for short ($n\le6$) alkane chains,
values taken from MP2 level calculations are sufficient to give $0.15$ kcal/mol
RMS accuracy\cite{gruzman2009} (taking CCSD(T) values as the correlation
benchmark).  Even for more demanding structures such as transition states
(pentane\cite{martin2013} for example) an RMS of $0.2$ kcal/mol is quite
satisfactory for many purposes.  Post MP2 contributions to the correlation
energy at the CCSD level provide a noticable improvement for transition states,
though there is a tendency for 
it to overcompensate by $0.2-0.5$ kcal/mol for more strongly rotated hexane
conformers.  The additional correlation energy coming from perturbative triples
is consistent with calculations performed on systems with dispersion dominated
interactions.\cite{takatani2010,simova2013}

\begin{figure}[b]
\includegraphics[width=8.26cm]{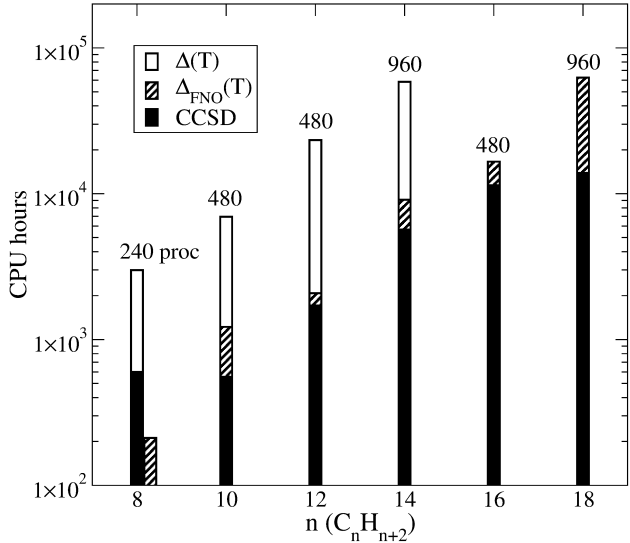}
\caption{\label{timing}Total CPU hours used to compute the correlation energy
for a single alkane conformer.  Also listed above each timing bar is the number
of processors used.  Calculations for the $n=8-12$ and $n=16$ chains were performed on the
HPC cluster while the $n=14$ and $n=18$ chain calculations were performed on the
new HiPerGator cluster.  The difference between clusters being the increased
number of avaliable processors and inter-node communication bandwidth.} 
\end{figure}

With our available computing resources we are able to compute accurate {\it ab
initio} coupled cluster conformer energies (CCSD(T)/cc-pVTZ) for the alkane
chains $n=8-14$, requiring 25,000 CPU hours for each $n=12$ conformer and
65,000 CPU hours for each $n=14$ conformer (see Figure \ref{timing}.  Obtaining
accurate values for this wide range of chain lengths allows us to benchmark any
further approximate methods that we choose to use in order to extend our
analysis to longer alkane chains.  The breakdown of the post SCF correlation
contributions through CCSD(T) for these medium length alkanes can be found in
Table \ref{benchE}.  Two systematic trends can be noticed here: the MP2
conformer energy consistently underestimates by $0.5$ kcal/mol while the CCSD
energy overshoots by $0.5$ kcal/mol, causing the correct value to fall directly
in between the two values.\cite{pitonak2009}  This illustrates the size of both infinite order
singles and doubles, providing a much more complete dispersion contribution, and
the significant role of connected triples in these extended systems.  

As mentioned earlier, while we are able to perform CCSD energy calculations
for all the chain lengths under consideration, the
$O(o^3v^4)$ cost of the perturbative triples becomes untenable for chains longer
than $n=14$ on available computational resources.  Therefore we
opted for an approximate treatment using the CCSD FNO(T) virtual space
truncation scheme.\cite{taube2005,taube2008}
Numerical tests show that an occupation number threshold of $1\times 10^{-4}$ is
sufficient to drop approximately 40\% of the virtual space while retaining
$\sim0.1$ kcal/mol accuracy compared to the full virtual space result, as
illustrated in Table \ref{fnoE}.  Because of the size extensivity of these
many-body methods this error grows very slowly with the number of electrons,
reaching $0.15$ kcal/mol at chain lengths of $n=14$.  A beneficial cancelation
of error can be observed here where the FNO composite method predicts conformer
energies between that of the CCSD(T) value and the 
CCSD(T)/CBS extrapolated CBS estimate.  Because of
computational cost considerations we include this $\pm 0.2$ kcal/mol variance
within the error estimate of the relative conformer energies. 

\begin{figure}[t]
\includegraphics[width=8.25cm]{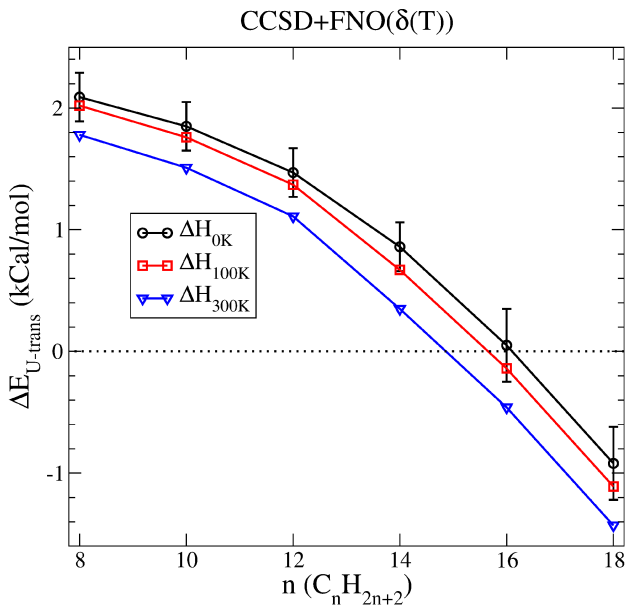}
\caption{\label{zpeshift}Calculated enthalpy differences, 
$\Delta H= \Delta H_{\rm hairpin}-\Delta H_{\rm linear}$, using
CCSD FNO(T)/cc-pVTZ single point energies and MP2/cc-pVTZ harmonic vibrational
frequencies including temperature dependent shifts.  }
\end{figure}

We compute the harmonic ZPE of the shorter alkane chains ($n=8-14$) at the
all-electron MP2 level of theory,
with the results presented in Table \ref{benchE}.  Thermodynamic considerations
are taken into account by computing the change in internal energy (vibration,
rotation and translation) as a function of temperature.  The increase in number
of degrees of freedom as the chains lengthen significantly increases the
computational cost, necessitating an approximate ZPE shift for the longer
alkanes.  Our final 
CCSD FNO(T)
relative conformer energies with ZPE
and temperature dependent shifts (simply referred to as $\Delta H$) are given in
Table \ref{fnoE}.  Noting the nearly constant ZPE and temperature shift in
$\Delta H$ for each of the shorter alkanes, we take as an approximation
that the $\Delta H$ shift for the longer ($n>14$) alkanes is the same as for the
$n=14$ alkane.  Because of this added approximation to the final relative
conformer energy we increase the estimated error bars to $0.3$ kcal/mol for the
longer alkane chains.  Our final $\Delta H$ values (plotted in Figure
\ref{zpeshift}) show that the $n=18$ alkane chain the hairpin conformer is
definitely preferred over the linear structure by a full kcal/mol.
For low temperatures ($0\sim100$ K) the $n=16$ hairpin conformer is possibly
preferred with the estimated error bars extending on either side of the 0 line,
however as the temperature increases our $\Delta H_{300K}$ value strongly
suggests that the $n=16$ hairpin is preferred.  These results are in complete
agreement with the gas jet experimental (performed at 100K) work of
L\"{u}ttschwager {\it et al.}\cite{luttschwager2013} where the hairpin
preference is found to be between $n=16$ and $18$.  

\begin{figure}[b]
\includegraphics[width=8.25cm]{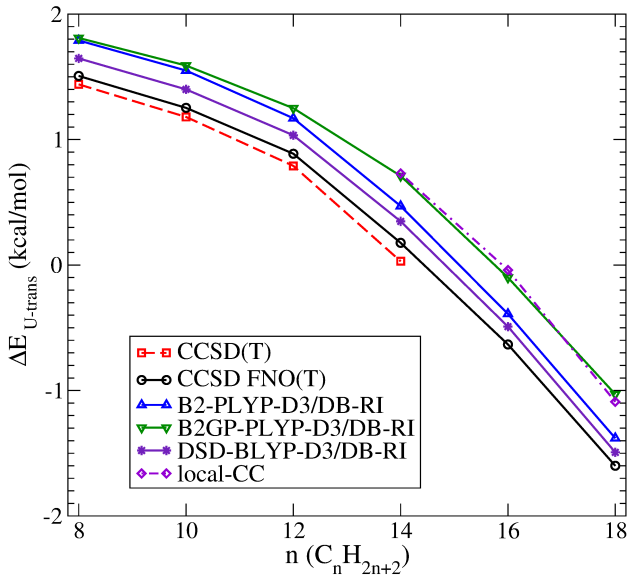}
\caption{\label{approximate}Composite relative single point energies,
$E_{\rm hairpin}-E_{\rm linear}$, compared against benchmark 
CCSD(T)/cc-pVTZ values.}
\end{figure}

With the FNO coupled cluster calculations costing 12\% of the corresponding full
virtual space calculations, we were able to perform 
CCSD FNO(T)
calculations for all the alkane chains through $n=18$.  Even so, the $O(o^3v^4)$
scaling remains such that the $n=18$ chain required 60,000 CPU hours for each
conformer to obtain the relative energy.  Clearly the attractiveness of an approximate
theory with a reduced computational scaling is great.  Qualitatively some
force field and semi-empirical calculations perform well, with
OPLS-AA\cite{thomas2006} and MM2\cite{goodman1997} both predicting that the
$n=18$ conformer energetically prefers the hairpin with PM3 predicting $n=12$
(this is excluding ZPE and other thermodynamic shifts).  However other commonly
used force field methods do not perform as well such as MM3 and AMBER,
which predict\cite{goodman1997} the hairpin turning point at $n=25$ and $n=26$
respectively.  Taking advantage of the polymer like repetitive structure of the
long alkanes L\"{u}ttschwager {\it et al.} have performed a composite local
coupled cluster (local-CC) correlation calculation for the $n=14-22$
series, including ZPE and thermal shifts, and compute that $n=18$ is the first
lowest energy hairpin length.  Qualitatively this is in good agreement with our own results,
though quantitatively we find the local-CC results to be too high compared to
our CCSD FNO(T) values, but very consistent with the DH-DFT results (see Table
\ref{totalsE} and Figure \ref{approximate}).  

\begin{table*}
\caption{\label{totalsE}{\rm Relative conformer energies
(hairpin $-$ linear, in kcal/mol) for the
$n=8-18$ alkane chains computed using various composite methods with the
cc-pVTZ basis set.  Also shown are the local-CC values from L\"{u}ttschwager
{\it et al.}\cite{luttschwager2013}  Benchmark CCSD(T) energies are included for
avaliable chain lengths.  Here * denotes the reported hairpin/linear crossing.
}}
\begin{tabular}{lrrrrrr}
Method
 & C$_{8}$H$_{18}$
 & C$_{10}$H$_{22}$
 & C$_{12}$H$_{26}$
 & C$_{14}$H$_{30}$ 
 & C$_{16}$H$_{34}$ 
 & C$_{18}$H$_{38}$ \\
\hline
CCSD(T)            & 1.44 & 1.18 & 0.79 & 0.03 \\
CCSD FNO(T) & 1.51 & 1.25 & 0.89 & 0.18 & -0.63 & -1.60 \\
B2-PLYP-D3/DB-RI & 1.79 & 1.55 & 1.17 & 0.47 & -0.39 & -1.38 \\
B2GP-PLYP-D3/DB-RI & 1.81 & 1.59 & 1.25 & 0.71 & -0.10 & -1.03 \\
DSD-BLYP-D3/DB-RI & 1.65 & 1.40 & 1.03 & 0.35 & -0.49 & -1.49 \\
local-CC\cite{luttschwager2013} &      &      &      & 0.73 & -0.04 & -1.09 \\
OPLS-AA\cite{thomas2006}
    &      &      &      &      &  2.2[-1]  &  1.3[-1] \\
MM2\cite{goodman1997}
    &      &      &      &      &       & * \\
PM3\cite{goodman1997}
    &      &      &  *   &      &       &   \\
\end{tabular}
\end{table*}

Density functional theory is not typically a good choice for non-covalently
bonded and weakly interacting conformer studies (or any system where dispersion
is an important consideration).  However with the inclusion of London type
dispersion\cite{hirschfelder1954} through an empirically derived additive
correction (in this case using Grimme's -D3 function\cite{grimme2010}) DFT
methods can be
quantitative\cite{gruzman2009,goerigk2010,goerigk2011,podeszwa2012} within the
limits of the training set.  In our previous work\cite{byrd2013-b} the accuracy
of the B2-PLYP\cite{grimme2006a}, B2GP-PLYP\cite{karton2008} and
DSD-BLYP\cite{kozuch2010} double-hybrid DFT composite functionals using the
dual-basis\cite{liang2004} resolution-of-the-identity\cite{katouda2009}
approximation (see the previous methods section) were tested for a variety of
non-covalently bonded dimers and small alkane conformers (so called
S22\cite{jurecka2006} and ACONF\cite{gruzman2009} test set).
The results of which were promising, with an average RMS error of $0.5$ kcal/mol for the
S22 set and $0.1$ kcal/mol for the $n=4-7$ alkanes.  Using these DH-DFT/DB-RI-MP2
composite methods, we return to the hairpin preference problem at hand and
compute relative conformer energies for all the long alkanes ($n=8-18$)
and compare the resulting values to our 
CCSD FNO(T) values in 
Table \ref{totalsE}.  As can be seen in Figure \ref{approximate} the DH-DFT/DB-RI-MP2
conformer energies track the higher level {\it ab initio} results very well, the
best curve coming from the DSD-BLYP functional with a nearly constant $0.1$
kcal/mol error.  With a computational cost many orders less than the predictive
CCSD FNO(T) composite method considered here, these DH-DFT/DB-RI-MP2
composite methods continue to be a promising approximate method in cases too
computationally difficult for coupled cluster calculations.

\section{Summary and Conclusions}

As the length of unbranched alkane chains reaches some critical length,
intramolecular dispersion forces cause a self-solvation effect in which the
chains assume a folded conformation.  To accurately determine this critical
chain length, linear and hairpin alkane conformer structures were optimized
using the MP2/cc-pVTZ level of theory for chains of length up through $n=18$.
Benchmark CCSD(T)/cc-pVTZ single point energy calculations were then performed
for octane through tetradecane using the ACES III\cite{acesiii2008} massively
parallel quantum chemistry package.  Harmonic zero point energies and
temperature shifts were computed using the MP2/cc-pVTZ level of theory.

For chains longer than $n=14$ it was necessary to use more approximate methods
to obtain conformer energy differences.  It was found that our
CCSD FNO(T) method which takes the full CCSD correlation energy and
adds the perturbative triples correlation energy taken from a frozen natural
orbital calculation where retaining only 60\% of the virtual space is required
to obtain results comparable to full CCSD(T) results (Table \ref{fnoE}).  We
have also explored the effectiveness of the dual-basis
resolution-of-the-identity double-hybrid density functional theory approach to
this problem.  With a computational cost several orders of magnitude less than
the more rigorous {\it ab initio} methods considered here we find that these
approximate DFT methods performed well with errors $\sim 0.2$ kcal/mol.

Computing the conformer temperature enthalpy differences using the
CCSD FNO(T) electronic energies and approximate MP2/cc-pVTZ ZPE shift
for alkane chains up through $n=18$ show that the temperature dependent hairpin
preference takes place at $n\ge 16$, with a confidence of $\sim 0.3$ kcal/mol.
This finding is in complete agreement with the experimental results of
L\"{u}ttschwagger {\it et al.}\cite{luttschwager2013}

As the temperature increases, entropic effects will become important.  The
Gibbs free energy difference between linear and folded chains will tend to
decrease due to the increased entropy of the folded conformation.  An accurate
assessment of this effect would require conformational sampling of the
CCSD(T) energy surface, a calculation that greatly exceeds the computational
resources available to us. However, the close agreement of our calculated
results with the low temperature results of 
L\"{u}ttschwagger {\it et al.}\cite{luttschwager2013}
suggests that our conclusions would be significantly altered by
entropic effects only at much higher temperatures.

\section{Acknowledgements}

JNB and RJB would like to acknowledge funding support from the United States
Army Research Office grant number W911NF-12-1-0143 and ARO DURUP grant number
W911-12-1-0365.

%


%


\end{document}